\documentclass[twocolumn,pra,aps,superscriptaddress]{revtex4}
\usepackage{amsmath,amsfonts}
\usepackage{graphicx}
\usepackage[bookmarks=true, colorlinks, linkcolor=blue, urlcolor=blue, citecolor=blue, plainpages=false, pdfpagelabels, final, breaklinks=true]{hyperref}

\begin{document}

\newcommand\bra[1]{\mathinner{\langle{\textstyle#1}\rvert}}
\newcommand\ket[1]{\mathinner{\lvert{\textstyle#1}\rangle}}
\newcommand\braket[1]{\mathinner{\langle{\textstyle#1}\rangle}}
\newcommand\Tr{\operatorname{{Tr}}}
\newcommand\hatc{\hat{c}}
\newcommand\hatf{\hat{f}}
\newcommand\calE{\mathcal{E}}
\newcommand\msc[1]{\textcolor{red}{#1}}

\title{Universal Entanglement Revival of Topological Origin}

\author{Dongni Chen}
\affiliation{Research Center for Advanced Optics and Photoelectronics, Department of Physics, College of Science, Shantou University, Shantou 515063, China}
\affiliation{Department of Physics, Korea University, Seoul 02841, South Korea}

\author{Stefano Chesi}
\affiliation{Beijing Computational Science Research Center, Beijing 100193, People's Republic of China}
\affiliation{Department of Physics, Beijing Normal University, Beijing 100875, People's Republic of China}

\author{Mahn-Soo Choi}
\email{choims@korea.ac.kr}
\affiliation{Department of Physics, Korea University, Seoul 02841, South Korea}

\begin{abstract}
We investigate the dynamics of entanglement in dissipative fermionic and bosonic Su–Schrieffer–Heeger (SSH) models and discover that they exhibit a revival dynamics of entanglement when the decoherence channel preserves the chiral symmetry.
This behavior is only observable in the topological phase, and the visibility of the revival diminishes to zero at the phase boundary.
Furthermore, the revival acquires a universal character, meaning that the entire time-evolution profile remains independent of system size, provided that the size exceeds the localization length of the edge modes associated with the topological phase.
Our findings suggest that the universal entanglement revival originates from the topological properties of the SSH model.
These dynamical properties may be experimentally accessible, for instance, by utilizing photonic quantum computers.
\end{abstract}

\maketitle

\section{Introduction}

Quantum entanglement, a critical resource in quantum science and technology, is fragile and prone to rapid degradation under quantum decoherence~\cite{Zurek2003_RevModPhys,Schlosshauer2007}. Consequently, systems exhibiting entanglement revival phenomena, where quantum correlations reemerge following initial decay despite persistent decoherence, have attracted significant interest~\cite{Zyczkowski2002_PRA,RTanas2004_JPB}.  
Ever since, this revival behavior has been extensively investigated across diverse quantum platforms, spanning both discrete- and continuous-variable systems~\cite{Benatti2006_JPA, Ficek2006_PRA, Cunha2007_NJP, Paz2008_PRL, Drumond2009_JPA, Das2009_JPB, Orszag2010_AOP, Mirko2018_PRA, Aolita2015_RPP}. Studies have demonstrated even more striking dynamical features, including entanglement sudden death and rebirth phenomena~\cite{Ficek2008_PRA, AbdelAty2008_JPB, Lopez2008_PRL, Aolita2015_RPP, Lakhfif2022_PLA, Nunavat2023_MPLA}.  
Such entanglement revivals must be distinguished from the direct influence of interaction which, by its very nature, also tends to enhance entanglement. The latter acts on a time scale given by the inverse interaction strength, whereas the entanglement revivals happen over a much longer time scale (typically, longer than the quantum decoherence time). 

Various physical mechanisms have been suggested to understand the occurrence of entanglement revival behavior, which usually involves a non-trivial interplay between coherent interactions and environmental decoherence~\cite{Zyczkowski2002_PRA,RTanas2004_JPB,Benatti2006_JPA,Ficek2006_PRA,Cunha2007_NJP,Paz2008_PRL,Drumond2009_JPA,Das2009_JPB,Orszag2010_AOP,Mirko2018_PRA,Aolita2015_RPP,Ficek2008_PRA,AbdelAty2008_JPB,Lopez2008_PRL,Lakhfif2022_PLA,Nunavat2023_MPLA,Chen2024_NJP}.
Of our particular interest is a generic scenario based on the \emph{chiral symmetry}, recently discussed for an ensemble of three-level systems interacting with single-mode bosons~\cite{Chen2024_NJP}. 
In this scenario, the chiral symmetry generates a distinct zero-energy state that demonstrates remarkable resilience against decoherence. Initially, the system experiences a phase where dissipation rapidly diminishes the initial entanglement. Subsequently, the system undergoes a gradual transition to the chiral symmetry-protected zero-energy state, eventually undergoing a self-purification process that entails a distinct revival of entanglement.

Chiral symmetry plays an especially prominent role in symmetry-protected \emph{topological} states of matter~\cite{Chiu2016_RevModPhys}. In this context, the associated zero-energy states become localized at the system’s edges and, according to the bulk-boundary correspondence principle, serve as distinctive signatures of topological phases.
Entanglement is both a fundamental quantum resource and a valuable tool for characterizing quantum phase transitions~\cite{Osterloh02a,Amico08a} as well as topological phase transitions~\cite{Kitaev06a,Fendley07b}.
Therefore, by combining these perspectives, it is natural to expect that the entanglement revival dynamics can provide valuable insights into the characterization of topological phases protected by chiral symmetry.

Motivated by this idea, in the present work, we investigate entanglement dynamics in a dissipative variant of the Su-Schrieffer-Heeger (SSH) model.
The SSH model represents one of the best-known examples of chiral symmetry-protected topological insulators~\cite{Su1979_PRL,JanosK2016,shen2017}, and has considerable relevance for quantum technology applications~\cite{Kitaev2003_AnnalsofPhysics,Freedman2003_BAMS,Nayak2008_RevModPhys, Bomantara2018_PRB,Bomantara2018_PRL,Ville2017_SciPostPhys}, due to its simple form and robustness against disorder~\cite{Halperin1982_PRB,Buttiker1988_PRB,Niu1984_JPA,Hasan2010_RevModPhys,Qi2011_RevModPhys,Ryu2010_NJP,Kawasaki2022_PRB}.
Recent studies have extended topological analysis to dissipative dynamics~\cite{Huang2020_PRB, Diehl2011_Nature,Bardyn2013_NJP,Rivas2013_PRB,Viyuela2012_PRB,Dangel2018_PRA,salatino2025}, yet the interplay between topological properties and entanglement dynamics remains underexplored. Understanding this relationship is critical for implementing topological systems in practical quantum technologies.
Our work systematically explores how the topological phases manifest through entanglement revival behaviors in the SSH model, establishing a characterization of topological phases in terms of entanglement dynamics.

We find that, in the topological phase of the SSH model, the dissipative entanglement dynamics not only exhibits a revival behavior but also takes a universal form;
in other words, the entire time-evolution profile of entanglement is independent of the chain length, with curves for different system sizes collapsing onto a single curve.
The universal profile becomes evident for chain lengths exceeding the localization length of the zero-energy edge modes, which are characteristic of the topological phase according to the bulk-boundary correspondence~\cite{Chiu2016_RevModPhys}.
In the trivial phase, however, all these features disappear.
All these suggest that the entanglement revival behavior is directly tied to the presence of the zero-energy edge modes, i.e., the topological state of the SSH model.

Our findings apply to both fermionic and bosonic versions of the SSH model, thus they can be simulated in tunable quantum systems, such as photonic quantum computers~\cite{Madsen2022_Nature,Brien2009_NaturePhotonics,yuan2024arxiv}.
On a photonic quantum computer, nearest-neighbor hoppings of variable coupling strengths can be realized with tunable beam splitters.  
Additionally, the required photon loss pattern with lower and higher rates at even and odd sites, respectively, can be simulated by selective leakage of photons.
Therefore, we expect that the topologically protected entanglement revival reported in this work is achievable with current photonic technology. We also note that a fermionic SSH model with tunable couplings has been experimentally realized using donor-based quantum-dot arrays in silicon~\cite{Kiczynski2022}, making this another promising platform for testing our predictions.

This paper is organized as follows: In Section~\ref{model}, we introduce the SSH model, briefly reviewing its chiral symmetry and topological properties.
The theoretical methods we used are detailed in Section~\ref{Paper::sec:method}, including the semiclassical approach and different entanglement measures.
Section \ref{Paper::sec:result}  presents the main results, highlighting the universal entanglement revival observed in both fermionic and bosonic SSH models, as well as its connection to topological edge states.
Section \ref{Sec: Summary} concludes the paper.
To further support the arguments in the main text, four appendices are provided: 
Appendix~\ref{Appendix:A} demonstrates the universal revival behavior in the mutual information, as an alternative to entanglement. 
Appendix~\ref{Appendix:B} examines the von Neumann and R\'enyi entropies \emph{per site} (i.e., the entropy densities), which also exhibit a universal behavior. Appendices~\ref{Appendix:C} and~\ref{Appendix:D} investigate more realistic conditions, including spatially correlated noise and finite temperature, and confirm that the universal revival remains qualitatively robust.

\section{Model}
\label{model}

\begin{figure*}
\centering
\includegraphics[width=100mm]{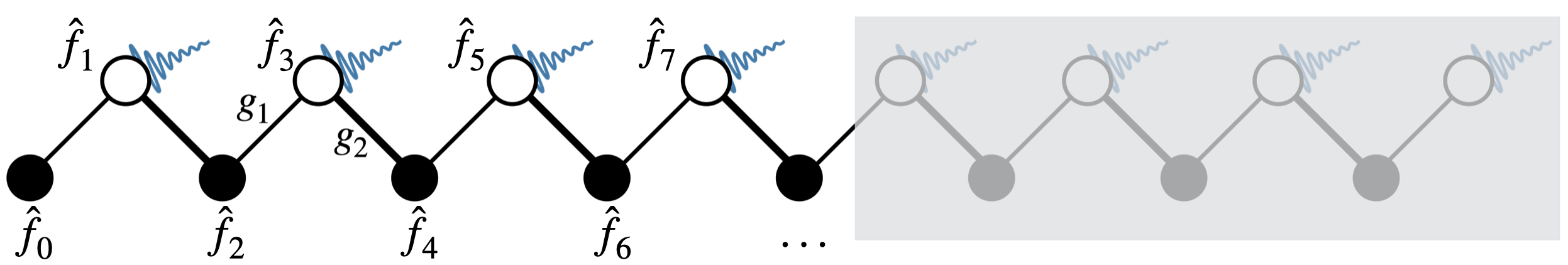}    
\caption{Schematics of the dissipative Su-Schrieffer-Heeger (SSH) model. The coupling strengths between nearest-neighbor sites alternate between two values, $g_1$ and $g_2$, and the quantum decoherence process (wavy lines) occurs only at the odd sites.
For numerical efficiency, we ignore the shaded region and refer to the remaining part as the `half-chain SSH mdoel'.}
\label{modelcp}
\end{figure*}

The SSH model is a prototype of one-dimensional (1D) topological insulators~\cite{JanosK2016,shen2017}. As a tight-binding model with alternating hopping amplitudes, $g_1$ and $g_2$, it exhibits a \emph{chiral symmetry} (also known as \emph{sub-lattice symmetry}), which, together with the time-reversal symmetry, completely classifies the topological states of matter. The chiral symmetry is characterized by a unitary operator that \emph{anti}-commutes with the Hamiltonian, and ensures that the Hamiltonian matrix can be written in a block-off-diagonal form, leading to a degenerate zero-energy subspace and a symmetric energy spectrum around the zero energy.
The zero-energy modes are especially interesting: According to the bulk-boundary correspondence principle, they always accompany the topologically non-trivial phase of topological insulators and are localized at the boundaries of the system.
In the SSH model, the topological phase occurs when $g_1<g_2$~\footnote{Without loss
of generality, we assume real and positive hopping amplitudes. Complex hopping amplitudes can always be brought to real positive values with a gauge transformation.}. In this phase, there are two zero-energy edge modes at the two ends of the chain. The \emph{localization length} of each edge mode,
\begin{equation}
\label{Paper::eq:locLength}
\xi = 1+\frac{2}{\log(g_2/g_1)},
\end{equation}
describes how far the mode spreads from the edge into the bulk of the chain.
For systems of sufficiently large size $L\gg\xi$, the zero-energy edge modes at the two ends are decoupled from each other. 
In this case, as long as any physical effects due to the zero-energy edge modes are concerned,
one can focus on each edge mode separately.
Therefore, for the sake of numerical efficiency, we will consider only half of the physically realistic chain, ignoring the shaded portion in
Fig.~\ref{modelcp}~\footnote{In this half-chain SSH model, one of the unit cells is not complete, as it contains only a single site instead of two, and the bulk-boundary correspondence does not hold in the usual sense.}.
This simplification allows us to concentrate on the properties and behavior of a single edge mode, while maintaining the essential features of the topological phase.
This half-chain SSH model contains $n$ two-site unit cells and a single unpaired site, so the overall chain length $L = 2n + 1$ is always odd. The Hamiltonian reads as
\begin{equation}
\label{Eq: Hmiltonian}
\hat H =
\sum_{i=1}^{(L-1)/2}\left( g_1\hat f_{2i-2}^\dagger \hat f_{2i-1}
+ g_2 \hat f_{2i-1}^\dagger \hat f_{2i}\right)+\mathrm{H.c.} ,
\end{equation}
where $\hatf_k^\dag$ and $\hatf_i$ are the creation and annihilation operators, respectively, at site $i = 0,1,\ldots , L-1$. 
In the fermionic (bosonic) SSH model, they satisfy the anti-commutation (commutation) relations,
\begin{math}
\{\hatf_i,\hatf_j^\dag\} = \delta_{ij}
\end{math}
\begin{math}
([\hatf_i,\hatf_j^\dag]=\delta_{ij}).
\end{math}
To better reveal the consequences of the chiral symmetry and the accompanying topological properties of the system, we rewrite the Hamiltonian in terms of the normal modes $\hat{c}_{\pm k}$:
\begin{equation}
\label{eq:HamiltonianNormal}
\hat{H} = \sum_{k=1}^{(L-1)/2}\epsilon_k
\left(\hat{c}_k^\dagger \hat{c}_k- \hat{c}_{-k}^\dagger \hat{c}_{-k}\right),
\end{equation}
where the single-particle energies $\pm\epsilon_k$ are given by:
\begin{equation}
\label{Eq: energy}
\epsilon_k := \sqrt{g_1^2 + g_2^2 + 2g_1 g_2 \cos(\tilde{k})},
\end{equation}
with
\begin{math}
\tilde{k} := 2\pi k/(L+1).
\end{math}
The normal modes are related to the bare fermion/boson modes $\hat{f}_i$ by
\begin{equation}
\hat{c}_{\pm k} 
= \sum_{i=0}^{(L-1)/2}A_{ki}\hat{f}_{2i} \pm \sum_{i=1}^{(L-1)/2}B_{ki}\hat{f}_{2i-1},
\end{equation}
with
\begin{subequations}
\label{Paper::eq:transform}
\begin{align}
\label{Paper::eq:transform:a}
A_{ki} & := \sqrt{\frac{2}{L+1}}\sin\left(\tilde{k} i + \phi_k\right), \\
\label{Paper::eq:transform:b}
B_{ki} & := \sqrt{\frac{2}{L+1}}\sin\left(\tilde{k} i\right) ,
\end{align}
\end{subequations}
and
\begin{math}
\cos(\phi_k) := (g_1 \cos\tilde{k} + g_2) / \epsilon_k.
\end{math}
Furthermore, there is a special normal mode which does not appear in Eq.~\eqref{eq:HamiltonianNormal} as it has the zero energy:
\begin{equation}
\label{Eq:edge}
\hat{c}_0 = \sqrt{\frac{1-(g_1/g_2)^2}{1-(g_1/g_2)^{L+1}}}
\sum_{i=0}^{(L-1)/2}(-g_1/g_2)^{i}\hat f_{2i}.
\end{equation}
Clearly, when $g_1<g_2$, this zero-energy mode has its probability amplitudes localized at the left end of the system \footnote{In the full SSH model, there is another zero-energy edge mode at the right end.}.
For later use, when $k=0$ we define:
\begin{equation}
\label{Paper::eq:transform:c}
A_{0i} = \sqrt{\frac{1-(g_1/g_2)^2}{1-(g_1/g_2)^{L+1}}}(-g_1/g_2)^{i}.
\end{equation}
Recall that, unlike this zero-energy edge mode, all other modes $\hat{c}_{\pm k}$ ($k\neq 0$) are bulk modes with probability amplitudes spread over the whole chain.

We are interested in the topological influence on the dynamics of quantum entanglement in the presence of decoherence.
To clearly illustrate the topological effects, we assume that the quantum decoherence processes preserve the chiral symmetry; that is, we assume that decoherence processes occur only on the odd sites, ${2i-1}$, with a uniform decay rate $\Gamma$. 
The noisy dynamics is governed by the quantum master equation for the density operator $\hat\rho$
\begin{equation}
\label{Paper::eq:master}
\frac{d\hat \rho}{dt} = -i[\hat H,\hat \rho]
+ \Gamma\sum_{i=1}^{(L-1)/2}\mathcal{D}[\hat f_{2i-1}]\hat \rho,
\end{equation}
where  $\mathcal{D}$ is the superoperator
\begin{equation}
\mathcal{D}[\hat f]\hat \rho
:= \hat f\hat \rho\hat f^\dagger
- \frac{1}{2}(\hat f^\dagger\hat f\hat\rho+\hat \rho\hat f^\dagger\hat f),
\end{equation}
describing the single-particle decay process and $\Gamma$ is the decay rate.

We will consider an many-body initial state of the form:
\begin{equation}
|\psi_g\rangle = \hat{c}_0^\dagger
\prod_{k=1}^{(L-1)/2} \hat{c}_{-k}^\dagger\,
\ket{\ },
\label{Eq:ins}
\end{equation}
where $\ket{\ }$ is the vacuum state. 
For the fermionic SSH model, this initial state is one of the two degenerate many-body ground states with all the negative-energy modes and zero-energy mode singly occupied; hence, it is a natural choice.
In the bosonic model, occupation of all the negative and zero-energy normal modes can be achieved by external driving, as commonly done in experimental demonstrations of topological band structures, e.g., in photonic crystals~\cite{Lu2014_NaturePhotonics,Rechtsman2013_Nature}.
Since we study non-equilibrium dynamics, the resulting physical effects inevitably depend on the initial state. However, as long as the zero-energy edge mode is initially occupied, our main findings remain qualitatively valid for many other initial states. 
In particular, we have checked that a universal revival is found for an initial state where all the normal modes are singly occupied.
This confirms that the universal entanglement dynamics reported below is closely related to topological features.

\section{Methods}
\label{Paper::sec:method}

Before presenting in Sec.~\ref{Paper::sec:result} our  main results, here we describe the methods used to solve the quantum master equation, Eq.~\eqref{Paper::eq:master}, and to analyze the entanglement content of the solution.

\subsection{Semiclassical approach}
\label{SEC.IIIA}

The computational cost to exactly solve the quantum master equation increases exponentially with system size~\footnote{For non-interacting \emph{fermions} with a restricted class of quantum jump operators, a polynomial-time method is available; See, for example, Refs.~\cite{Prosen2008_NJP,Bravyi12a,Alba2023_SciPostPhys}.}.
Fortunately, in the parameter regime of our interest,
a semicassical approach provides an accurate description of the dynamics. We first note that, when the coupling strengths are much larger than the decay rate ($g_1,g_2\gg\Gamma$), the eigenstates of Eq.~\eqref{Eq: Hmiltonian} are generically well-separated in energy. 
Since the coherent phase factor between energy-separated eigenstates oscillates rapidly in time, it gets averaged out throughout the slow dissipative evolution.
We can therefore assume that $\hat\rho(t)$ is approximately diagonal in the many-body eigenstates, $\ket{E_\alpha}$:
\begin{equation}
\label{Paper::eq:semiclassicalRho}
\hat\rho(t)
\approx\sum_\alpha|E_\alpha\rangle P_\alpha (t)\langle E_\alpha|.
\end{equation}
The probabilities $P_\alpha$ for different many-body eigenstates $|E_\alpha\rangle$
satisfy the master equation
\begin{equation}
\label{Paper::eq:classicalME}
\frac{dP_\alpha(t)}{dt}=\sum_{\beta} \left ( \gamma_{\alpha\beta}P_\beta(t)- \gamma_{\beta\alpha}P_\alpha(t)\right),
\end{equation}
where the transition rate $\gamma_{\alpha\beta}$ from $\ket{E_\beta}$ to $\ket{E_\alpha}$  is given by
\begin{equation}
\gamma_{\alpha\beta} 
:=\Gamma\sum_{i=1}^{(L-1)/2} |\langle E_\alpha| \hat{f}_{2i-1} |E_\beta \rangle|^2.
\end{equation}
In terms of the normal-mode operators $\hatc_k$, it reads as
\begin{equation}\label{gamma_rate_general}
\gamma_{\alpha\beta} 
=\frac{\Gamma}{2} \sum_{k\neq 0} |\langle E_\alpha| \hat{c}_k |E_\beta \rangle|^2,
\end{equation}
which follows from the orthogonality of coefficients $B_{ki}$,
\begin{equation}
\label{Paper::eq:B:orthogonality}
\sum_{i=1}^{(L-1)/2}B_{ki}B_{k'i}= \delta_{kk'}/2.
\end{equation} 
In fact, this orthogonality relation has its origin in the chiral symmetry, leading to the \emph{separately} normalized coefficients $A_{ki}$ and $B_{ki}$; see Eqs.~\eqref{Paper::eq:transform} and \eqref{Paper::eq:transform:c}.
Therefore, we also have
\begin{equation}
\label{Paper::eq:A:orthogonality}
\sum_{i=0}^{(L-1)/2}A_{ki}A_{k'i} = \delta_{kk'}/2.
\end{equation}

We note some unique properties of the semiclassical dynamics described by Eqs.~\eqref{Paper::eq:semiclassicalRho} and \eqref{Paper::eq:classicalME}:
First, every time a decay process occurs, the total number of particles decreases by one. However, the zero-energy edge mode can never decay, because $\hat{c}_0$ has non-zero amplitudes only at the even sites, see Eq.~\eqref{Eq:edge}. 
In the long-time limit ($t\to\infty$), only one particle remains, occupying the zero-energy edge mode; i.e.,
\begin{equation}
\hat{\rho}_\infty := 
\lim_{t\rightarrow\infty}\hat{\rho}(t) = 
\hat{c}_0^\dagger\ket{\ }\bra{\ }\hat{c}_0.
\end{equation}
Second, with the initial condition in Eq.~(\ref{Eq:ins}), the occupation numbers of the normal modes are either one or zero, and a decay process corresponds to one of the occupied modes becoming empty. From Eq.~(\ref{gamma_rate_general}), we see that all these decays occur with an identical rate, $\Gamma/2$, which allows us to solve the semiclassical evolution of Eqs.~\eqref{Paper::eq:semiclassicalRho} and \eqref{Paper::eq:classicalME} in a closed form as
\begin{equation}
\label{Eq:rho}
\hat \rho(t)
\approx \sum_{m=0}^{(L-1)/2} P_{m+1}(t)\sum_{\{k_i\}_{i=1}^m}
\hat{c}_{\{k_i\}}^\dagger \hat\rho_{\infty} \hat{c}_{\{k_i\}},
\end{equation}
with
\begin{equation}
\label{Eq:probability}
P_{m+1}(t) 
= e^{-{m\Gamma t}/{2}}
\left(1 - e^{-{\Gamma t}/{2}}\right)^{\frac{L-1}{2}-m} .
\end{equation}
In Eq.~\eqref{Eq:rho}, we have used the short-hand notation
\begin{equation}
\hat{c}_{\{k_i\}_{i=1}^m} := \prod_{i=1}^{m} \hat{c}_{-k_i} ,
\end{equation}
where $\{k_i\}_{i=1}^m$ is an unordered set of mode indices (with $k_i >0$). 
We see from Eq.~(\ref{Eq:rho}) that the probabilities $P_\alpha$ for the many-body eigenstates $\alpha$ to host $m-1$ particles at negative energies (and one particle in the $\hat{c}_0$ mode) are all equal. 
For those states, we can simply set $P_\alpha =P_m$.

So far, we have justified the semiclassical  approximation in the large-coupling limit $g_1,g_2 \gg \Gamma$. However, the above description is more general, and allows us to compute the dynamics deep in the topological phase ($g_1 \ll g_2$), when $g_1$ can become comparable to $\Gamma$. In this limit, the energy differences between the eigenstates with the same $m$ appearing in Eq.~(\ref{Eq:rho}) are of order of $g_1$, thus the corresponding coherent phase factors would not oscillate on a fast timescale. 
Fortunately, however, the dissipative evolution does not generate such coherence, as can be checked by applying the full master equation, Eq.~(\ref{Paper::eq:master}), to a generic eigenstate $|E_\alpha \rangle\langle E_\alpha|$. 
This is due to a destructive interference of amplitudes, which follows from the orthogonality of the normal modes; i.e., Eqs.~\eqref{Paper::eq:B:orthogonality} and \eqref{Paper::eq:A:orthogonality}.
Nevertheless, it is still necessary to satisfy the condition $g_2 \gg \Gamma$
in order to prevent the full dynamics from generating coherence between many-body eigenstates with large energy differences of order $g_2$.

\begin{figure}
\centering
\includegraphics[width=80mm]{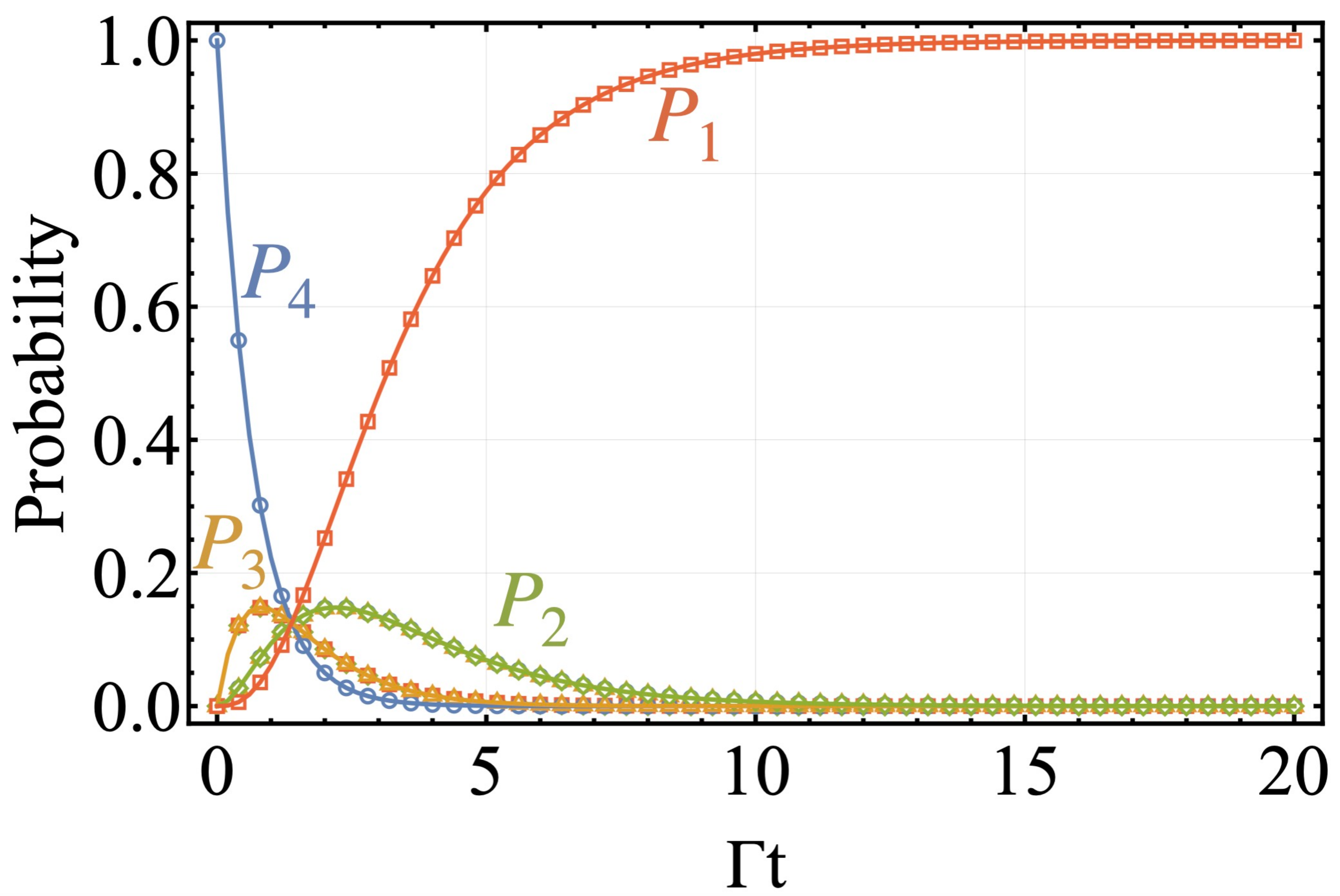} 
\caption{Probability $P_{m}$ for the system to be found in any many-body eigenstate with $m$ particles as a function of time. The empty markers are obtained by solving the quantum master equation whereas the solid lines are from the semiclassical description. We used $L=7$ and $g_1/g_2=\Gamma/g_2=0.1$.}  
\label{Fig:probability}
\end{figure}

To demonstrate the accuracy of the semiclassical approximation when $g_1\approx \Gamma$ (yet $g_2\gg\Gamma$), we compare in Fig.~\ref{Fig:probability} the numerical solution of the full quantum master equation~\eqref{Paper::eq:master} and the semiclassical approximation~\eqref{Eq:probability}.
Clearly, we observe an excellent agreement between the two, justifying the semiclassical approach.
The validity of the semiclassical approximation can be understood in terms of two complementary mechanisms. First we note that the coherence between eigenstates with the same number of particles (occupying the normal modes $\hat c_i$) vanishes due to destructive interference between decay paths into different eigenstates. This property can be derived from Eqs.~(\ref{Paper::eq:B:orthogonality}) and~(\ref{Paper::eq:A:orthogonality}). Specifically, when the system evolves from one initial state into a coherent superposition of such eigenstates, the amplitudes associated with different paths acquire opposite phases. These phase differences, which originate from the symmetry and spatial structure of the modes, lead to a cancellation of coherence. Furthermore, for eigenstates with different particle-numbers, coherence terms are suppressed by rapid oscillations arising from large energy gaps, set primarily by $g_2$. As long as $g_2 \gg \Gamma$, these coherence terms effectively average out.
In passing, we note that the maximum probability appears earlier for initial states with more particles, which is intuitively appealing. 

The semiclassical state in Eq.~\eqref{Eq:rho} provides a complete description of the dynamics. For fermionic systems, the state satisfies Wick's theorem~\cite{Mahan00a}, hence one can further rewrite it into a Gaussian form~\cite{Peschel2003_JPA}:
\begin{equation}
\label{Eq:Gassianrho}
\hat\rho = \left(\det G\right)
\exp\left[\sum_{i,j=0}^{L-1}K_{ij}(t)\hat f_i^\dagger \hat f_j\right],
\end{equation}
with the Gaussian kernel $K$ given by:
\begin{equation}
K = \ln\left[(1-G)G^{-1}\right].
\end{equation}
Here, $G$ denotes the matrix of single-particle Green's functions,
\begin{equation}
G_{ij}(t) 
:= \left\langle\hat f_i \hat f_j^\dagger\right\rangle 
= \Tr\left[\hat \rho(t)\hat f_i \hat f_j^\dagger\right],
\end{equation}
which can be conveniently obtained by first evaluating the Green's functions of the normal modes $\hat{c}_k$. Since the density matrix $\hat \rho$ in  Eq.~\eqref{Eq:rho} is diagonal in the normal modes,  we immediately find that
\begin{math}
\chi_{kk'} := \langle \hat{c}_k\hat{c}_{k'}^\dagger\rangle= \chi_{k}\delta_{kk'}.
\end{math} 
Explicitly, $\chi_k$ is obtained as follows:
\begin{equation}\label{chi}
\chi_{k}(t)=
\begin{cases}
1-e^{-\Gamma t/2} & (k<0),\\
0 & (k=0),\\
1 & (k>0).
\end{cases}
\end{equation}
Finally, $G$ is related to $\chi$ by the same canonical transformation between $\hat{f}_i$ and $\hat{c}_k$, given in Eqs.~\eqref{Paper::eq:transform} and \eqref{Paper::eq:transform:c}. By using Eq.~(\ref{chi}), it can be written as
\begin{equation}\label{G_matrix}
G_{ij}(t)=\sum_{k=1}^{(L-1)/2} \left[ (1-e^{-\Gamma t/2})U_{-k,i} U_{-k,j}+U_{k,i} U_{k,j}\right],
\end{equation}
where $U_{\pm k,2i}=A_{ki}$ and $U_{\pm k,2i-1}=\pm B_{ki}$.

The Gaussian form~\eqref{Eq:Gassianrho} of the density matrix turns out to be extremely powerful when later calculating the entanglement content.
We emphasize that for bosons, the mixed state in Eq.~\eqref{Eq:rho} is not Gaussian anymore, despite having exactly the same form.

\subsection{Entanglement measures}

Quantifying quantum entanglement~\cite{Plbnio2007_QInfoCompu} is vital in quantum information theory, yet is highly non-trivial to calculate, especially, for mixed states.
Common entanglement measures include entanglement entropy~\cite{Bennett1996_PRA}, concurrence~\cite{Wootters1998_PRL}, and logarithmic negativity (or the closely related negativity)~\cite{Plenio2005_PRL,Jens1999_JModopt}. However, entanglement entropy is applicable only to pure states, while concurrence is practical only for two-qubit systems, whether in pure or mixed states.
For mixed states of more general systems, the logarithmic negativity (or negativity) has been widely used as one of very few practically computable entanglement measures. Nevertheless, as it requires finding the eigenvalues of the density matrix after partial transposition, its computational cost increases quickly with system size.
To avoid complications arising from partial transposition, quantum mutual information is often used as well.  Although mutual information is not an entanglement measure in the strict mathematical sense, it still captures many important aspects of quantum correlations as well as classical correlations. Below, we will investigate the entanglement content of the mixed state~\eqref{Eq:rho} using the logarithmic negativity. However, as we illustrate in Appendix~\ref{Appendix:A}, our universal features are observed in the mutual information as well. 

The logarithmic negativity between two subsystems $\mathbb{A}$ and $\mathbb{B}$ is defined by~\cite{Plenio2005_PRL}
\begin{equation}
\label{Paper::eq:logneg}
\mathcal{E}(\mathbb{A},\mathbb{B}) := \log_2\left\|\hat\rho^{T_\mathbb{A}}\right\|
= \log_2\left\|\hat\rho^{T_\mathbb{B}}\right\|,
\end{equation}
where $\hat\rho^{T_\mathbb{A}}(\hat\rho^{T_\mathbb{B}})$ is the partial transpose of the density matrix $\hat\rho$ with respect to subsystem $\mathbb{A}$ ($\mathbb{B}$), and $\|\cdot\|$ denotes the trace norm.
As mentioned above and seen in the definition~\eqref{Paper::eq:logneg}, 
the main obstacle in computing the logarithmic negativity for large systems comes from the partial transposition:
If not for the partial transposition, in many cases the trace norm may be calculated without much difficulty, by using the symmetries of the system to decompose the large density matrix into smaller blocks. More specifically, in our model the total number of particles is conserved, i.e.,    
\begin{math}
[ \hat{n}_\mathbb{\mathbb{A}} + \hat{n}_\mathbb{\mathbb{B}}, \hat{\rho}] = 0 .
\end{math}
On the other hand, the partial transpose, $\hat\rho^{T_\mathbb{A}}$ or $\hat\rho^{T_\mathbb{B}}$, violates this symmetry, and one has to deal with the Hilbert space as a whole. Actually, in our model the difference $\hat{n}_\mathbb{A}-\hat{n}_\mathbb{B}$ in the number of particles is conserved~\cite{Qiu2022_CTP}, i.e,
\begin{math}
[\hat{n}_\mathbb{A} - \hat{n}_\mathbb{B}, \hat{\rho}^{T_\mathbb{A}}] = 0,
\end{math}
but the invariant blocks under this symmetry are still exponentially large.

Recently, a new type of logarithmic negativity based on a partial time-reversal transformation has been suggested for \emph{fermionic} systems~\cite{Shapourian2017_PRB, Shapourian2019_JSM,Shapourian2019_PRA}. 
Compared to the conventional (or bosonic) logarithmic negativity, this fermionic logarithmic negativity was shown to better capture the contributions to entanglement from Majorana fermions.
From our point of view, however, the key advantage lies in the fact that the partial time-reversal transformation of a fermionic Gaussian state remains Gaussian, allowing the fermionic logarithmic negativity to be calculated using a $2 L \times 2 L$ matrix of generalized Green's functions~\cite{Shapourian2017_PRB}; that is, with a polynomial amount of resources and time.
In our case, as the pairing correlations are zero, the fermionic logarithmic negativity is obtained directly from the $L\times L$ matrix of Green's functions, see Eq.~(\ref{G_matrix}), and we could study half-chain SSH chains up to $L\sim 10^3$ without difficulties.
For the bosonic case, on the other hand, the calculations are limited to relatively short chains.

\subsection{Fermionic vs bosonic SSH models}
 
Above, we have already made several remarks on technical differences between the bosonic and fermionic SSH models. Here, we summarize both fundamental and technical  differences between them. It is remarkable that, despite these differences, the fermionic and bosonic models share the same universal dynamics of entanglement.
 
Both bosonic and fermionic systems in the half-chain SSH model exhibit a chiral symmetry, characterized by the existence of a unitary operator that anti-commutes with the Hamiltonian. This symmetry leads to a degenerate zero-energy subspace and a symmetric single-particle energy spectrum around the zero energy. Additionally, the zero-energy edge mode reflects the non-trivial topological character of the band structure.
 
For large bosonic systems, the size of the Hilbert space poses significant computational challenges. Certain techniques, such as block diagonalization of $\hat \rho^{T_\mathbb{A}}$, are employed to alleviate the computational burden. In contrast, fermionic systems can benefit from the fact that the relevant state is Gaussian, since the partial time-reversal transformation of Eq.~(\ref{Eq:Gassianrho}) remains Gaussian.
 
In the next Section, we will present our results for both bosonic and fermionic systems in parallel, to highlight the common universal behavior despite the differences summarized above. 
We attribute the common features of entanglement dynamics to the chiral symmetry and topological properties, especially the zero-energy edge modes.

\section{Results}
\label{Paper::sec:result}

We now present our findings, which provide deeper insights into the relationship between dynamics and topological properties. 
Most of our calculations have been done using the symbolic quantum simulation framework Q3\cite{q3}.

\begin{figure}
\includegraphics[width=80mm]{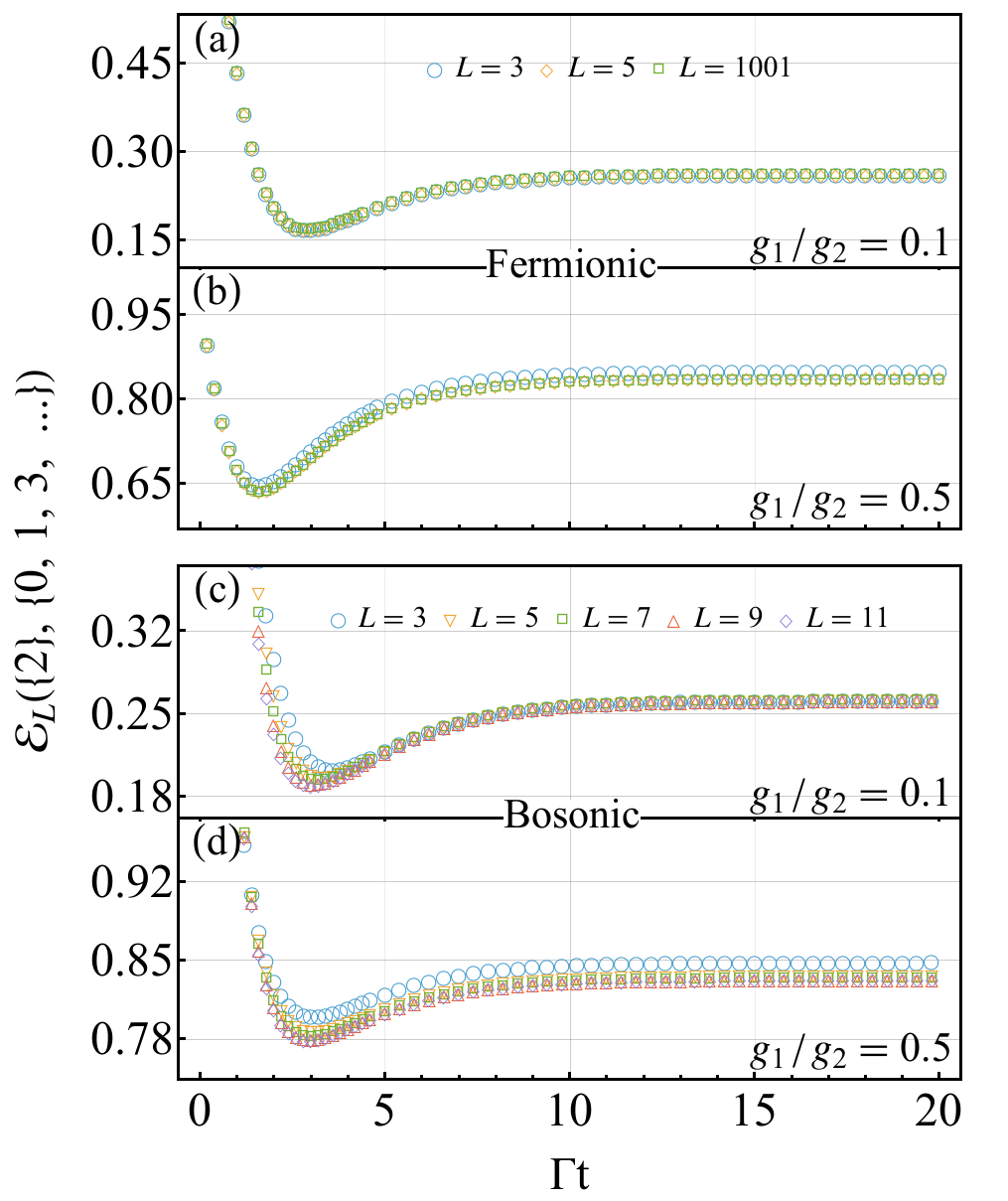} 
\caption{Universal entanglement revival, obtained by computing the logarithmic negativity $\mathcal{E}_L (\{2\}, \{0, 1, 3, \dots\})$ between the single-site subsystem $\{\hat f_2\}$ and the rest of the chain. Panels (a) and (b) consider the fermionic model at $g_1/g_2 = 0.1$ and $0.5$, respectively. Panels (c) and (d) are for the bosonic model, also using $g_1/g_2 =0.1$ and 0.5.  Other parameters are $g_2 = 1$ (arbitrary units) and $\Gamma/g_2 = 0.1$.}
\label{Fig:universalrevival}
\end{figure}

\subsection{Universal entanglement revival}

Figure~\ref{Fig:universalrevival} summarizes our main results, concerning the time evolution of entanglement between the second even site and the rest of the system (the qualitative features do not depend on a specific partitioning, see below).  
Figure~\ref{Fig:universalrevival} (a) and (b) refer to the fermionic half-chain SSH chain, where we computed the fermionic logarithmic negativity for two values of $g_1$ deep in the topological phase ($g_1 \ll g_2$). On the other hand, 
Fig.~\ref{Fig:universalrevival} (c) and (d) show the conventional logarithmic negativity of the bosonic model in the same phase.
In the fermionic case, we could calculate the fermionic logarithmic negativity for systems as large as $L\sim 10^3$; three representative sizes are shown in Fig.~\ref{Fig:universalrevival} (a) and (b).
In the bosonic case, the calculations were limited to relatively small values of $L$, because of the computational difficulties discussed in Section~\ref{Paper::sec:method}; the results for $L=3,5,\dots,11$ are shown in Fig.~\ref{Fig:universalrevival} (c) and (d).
In both cases, two distinctive features are clearly seen:
First, the entanglement revives after an initial decrease.
Considering the very fragile nature of quantum entanglement in the presence of decoherence, an entanglement revival is unusual and has attracted much interest~\cite{Chen2024_NJP,Zyczkowski2002_PRA,Aolita2015_RPP,RTanas2004_JPB,Benatti2006_JPA,Cunha2007_NJP,AbdelAty2008_JPB,Ficek2006_PRA,Ficek2008_PRA,Paz2008_PRL,Drumond2009_JPA,Das2009_JPB,Orszag2010_AOP,Nunavat2023_MPLA,Lakhfif2022_PLA,Mirko2018_PRA}.
Second, the entire temporal profile of the entanglement has a universal dependence, i.e., it is almost unaffected by the system size. 
Although we only show the entanglement between site $\hat f_{2}$ and the remaining sites $\{\hat f_0, \hat f_1, \hat f_3, \dots \hat f_L\}$ in Fig.~\ref{Fig:universalrevival}, a universal revival behavior can be observed regardless of the specific choice of partitioning, as long as one subsystem is confined to the neighbor of the left edge. 
However, the visibility of the revival does depend on the partitioning.


As demonstrated in a recent work~\cite{Chen2024_NJP}, the entanglement revival itself is due to a self-purification process directly related to the chiral symmetry and the associated zero-energy mode (not necessarily localized to the edges), which behaves as a dark state over the system’s evolution.  The system starts from a pure state with relatively large entanglement and, as it undergoes the decoherence process,  all the eigenmodes $\hat c_{k\neq 0}$ of Eq.~\eqref{Eq:ins} (which have nonzero amplitudes on the odd sites $\hat f_{2i-1}$) are subject to decay and decoherence.  
Consequently, the system evolves into a mixture of various intermediate states, resulting in a reduced purity and the suppression of entanglement. 
After this intermediate stage, however, the system begins to recover its purity (hence, its entanglement), because all particles have the fate of decaying out, except for those occupying the zero-energy mode, which has nonzero amplitudes only on the even sites.
Eventually, the system stabilizes to the final state, i.e., the zero-energy mode, which is immune to the decay process. The entanglement in the steady state is always determined by this zero-energy mode.

While the above arguments explain the presence of entanglement revival, its universal character is more intriguing, and is the main topic of the rest of the paper.
In the following, we perform a detailed analysis of the size dependence, revealing that the universal behavior is attributed to the topological feature of the system, especially the zero-energy edge mode.

\subsection{Topological effects through the edge mode}

\begin{figure}
\centering
\includegraphics[width=80mm]{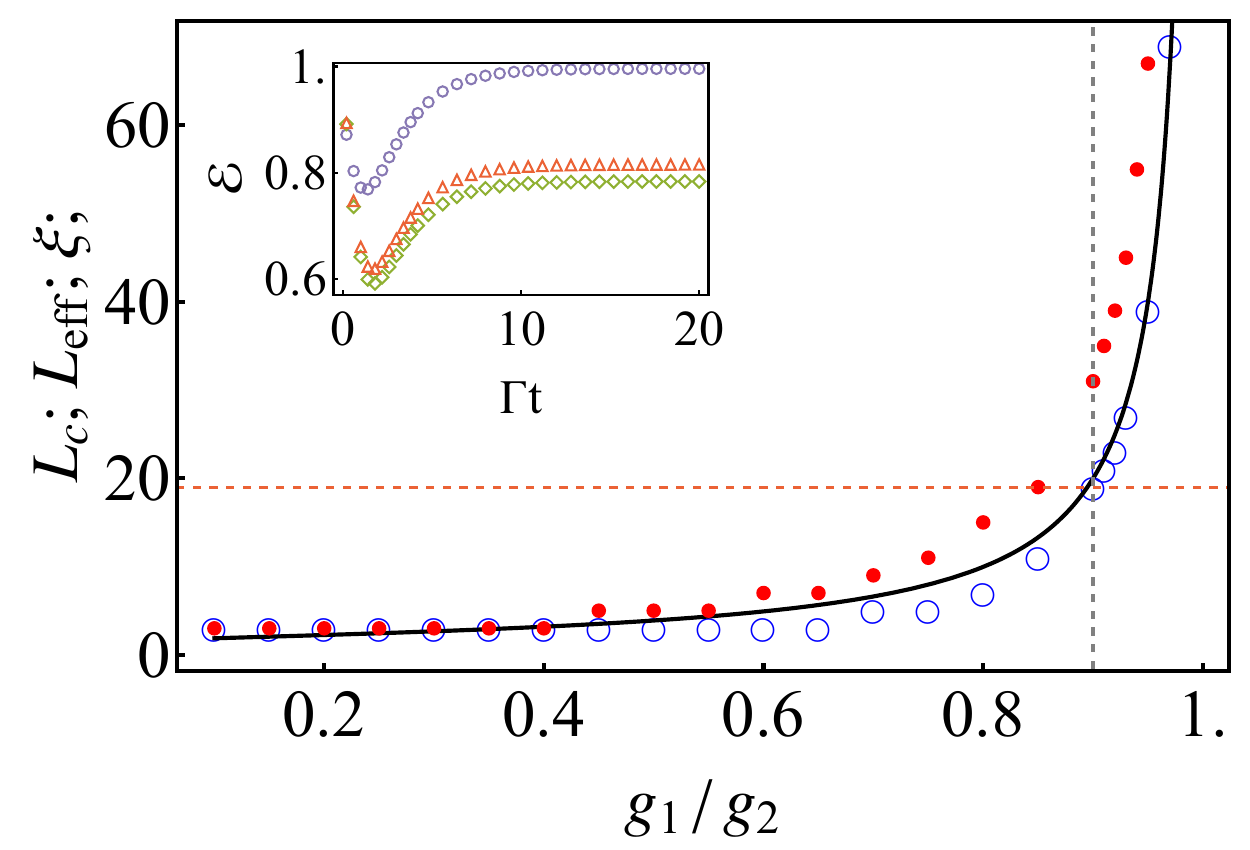}
\caption{
Comparison of the crossover length $L_\mathrm{cr}$ (blue empty circles) and the effective length $L_{\mathrm{eff}}$ (red filled circles) with the localization length $\xi$ of the edge mode (black solid line), as a function of the coupling ratio $g_1/g_2$.
The red dashed line marks the values $L_\mathrm{cr}\approx 19$, computed at $g_1/g_2=0.9$. Inset: time dependence of the logarithmic negativity between the single-site subsystem $\{\hatf_2\}$ and the rest of the chain. Here we take $g_1/g_2=0.9$ and $L=3$, $19$, $1001$ (from top to bottom).}  
\label{Fig:certainlength}
\end{figure}

In Fig.~\ref{Fig:universalrevival}, we have seen that the entanglement revival deep in the topological phase ($g_1/g_2\ll 1$) is universal, having a dynamical profile independent of system size.
On the other hand, as the boundary between the topological and trivial phases is approached ($g_1/g_2\to 1$), the evolution of entanglement shows significant variations especially for smaller systems.
When $g_1/g_2=0.9$, for example, the dynamical profile for $L=3$ is distinct from the profiles of larger systems ($L=19, 1001$), as shown in the inset of Fig.~\ref{Fig:certainlength}.
We find that the dynamical profiles remain universal as long as the system size is sufficiently large compared with a certain length scale, directly related to the localization length $\xi$ of the zero-energy edge mode; see Eq.~\eqref{Paper::eq:locLength}. 
In the following, we examine this crossover scale in detail.

We define two length scales concerning the entanglement revival dynamics:
The \emph{crossover length} $L_\mathrm{cr}$ is defined as the minimum system size which, at a given ratio $g_1/g_2$, achieves a universal time dependence. 
In other words, the dynamical profiles of systems larger than $L_\mathrm{cr}$ are all universal (within a certain numerical tolerance $\epsilon$).
More specifically, let $\mathcal{E}_L(t)$ be the logarithmic negativity for a system of length $L$. Then, choosing a sufficiently large time $T$, at which the entanglement has reached its asymptotic value, we require that for all $L\geq L_\mathrm{cr}$
\begin{equation}
\label{Eq:clength}
\frac{1}{T}\int_0^T{dt}\,\left|\calE_L(t)-\calE_\infty(t)\right| \leq
\epsilon\,\calE_\infty(T).
\end{equation}
Numerically, we approximate the $L\to \infty$ limit $\calE_\infty$ by taking $L=1001$. We also choose $T=20$ and $\epsilon=0.05$.
Figure~\ref{Fig:certainlength} plots the crossover length $L_\mathrm{cr}$ (blue empty circles) as a function of $g_1/g_2$. 
One can see that $L_\mathrm{cr}$ closely resembles the localization length $\xi$ of the edge mode (black solid line). 
For instance, when $g_1/g_2 = 0.9$, the crossover length matches the localization length $L_\mathrm{cr}\approx\xi\approx 19$. 
This value is highlighted by a horizontal dashed line in the main panel of Fig.~\ref{Fig:certainlength}. 
In the inset, the detailed entanglement profile with $L=19$ follows quite faithfully the universal dynamics at $L=1001$.

The second length scale we investigate is the \emph{effective length} $L_\mathrm{eff}$ of the system. To define $L_\mathrm{eff}$,
we consider the reduced density matrix 
\begin{math}
\hat\rho_j(t) := \Tr_{i=j+1}^L\hat\rho(t),
\end{math}
obtained by tracing out sites $j+1,\cdots,L$. In general, the reduced density matrix $\hat\rho_j(t)$ represents a physically different state from the full density matrix $\hat\rho(t)$.
However, in our case, it turns out that $\hat\rho_j(t)$ gives physically equivalent results to $\hat\rho(t)$, as long as $j$ is larger than a certain length scale, which we refer to as the effective length.
That is, the effective length $L_\mathrm{eff}$ is the minimum length at which the reduced density matrix has the same properties of the full density matrix,
$\hat\rho_j(t)\approx\hat\rho(t)$ for $j\geq L_\mathrm{eff}$.
The comparison between $\hat\rho_j(t)$ and $\hat\rho(t)$ is done in terms of the logarithmic negativity, in a manner analogous to Eq.~\eqref{Eq:clength}. More precisely, in Eq.~\eqref{Eq:clength} we substitute $\mathcal{E}_L(t) \to \mathcal{E}_j(t)$, where $\mathcal{E}_j(t)$ is the logarithmic negativity obtained from $\hat\rho_j(t)$. Ideally, we would compute $\hat\rho_j(t)$ from the full density matrix $\hat\rho(t)$ of the infinite half-chain SSH chain, i.e., in the limit $L\to \infty$. In practice, we can choose a sufficiently large value  of $L$ and, as before, we have used $L=1001$ in our numerical calculations. 

The explicit dependence of $L_\mathrm{eff}$ on $g_1/g_2$ is shown in Fig.~\ref{Fig:certainlength} (red filled circles). 
It is seen that like the crossover length $L_\mathrm{cr}$, the effective length $L_\mathrm{eff}$ also agrees well with the localization length $\xi$ of the edge mode. 
In short, both $L_\mathrm{cr}$ and $L_\mathrm{eff}$ follow closely the localization length $\xi$, and this strongly suggest that the spatial range of the edge mode is a critical factor in determining the onset of the universal entanglement revival behavior.

\begin{figure}
\centering
\includegraphics[width=80mm]{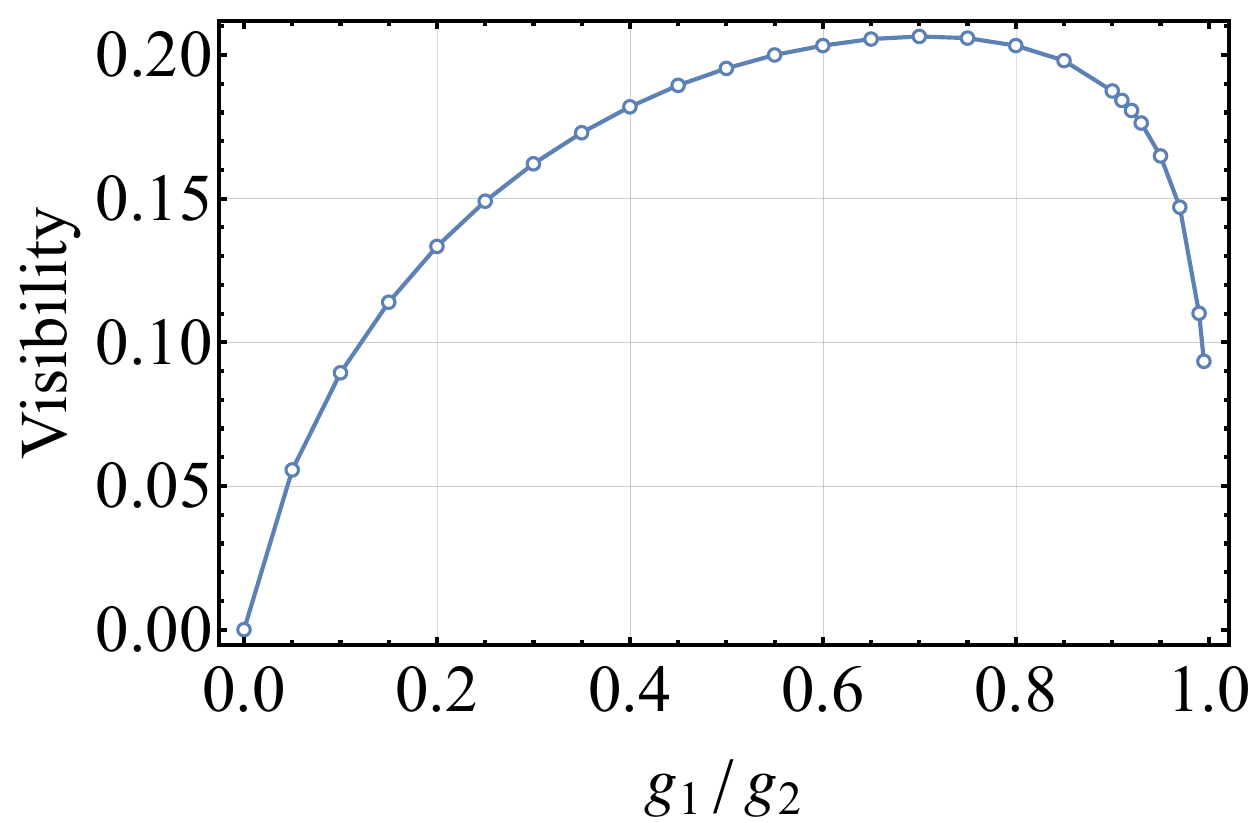}
\caption{The visibility of entanglement revival as a function of the coupling ratio $g_1/g_2$, for $L=1001$ and $\Gamma/g_2=0.1$.}
\label{Fig:visibility}
\end{figure}

Another feature that supports the topological origin of the entanglement revival is the visibility, 
defined as the difference between the asymptotic value ($t\to\infty$) and the minimum of the logarithmic negativity.
The details of the visibility differ between the fermionic and bosonic cases, and are affected by system parameters as well. 
However, as illustrated in Fig.~\ref{Fig:visibility} for a long fermionic chain,
we find the following generic qualitative features: The visibility is zero at $g_1/g_2=0$ and first increases with $g_1/g_2$, reaching a maximum around $g_1/g_2\approx 0.7$. Then, it sharply drops to zero (i.e., the entanglement revival disappears) as $g_1/g_2$ approaches 1.
At $g_1/g_2=0$, the visibility vanishes for a trivial reason, i.e., because the unit cells are decoupled.
However, the disappearance of the entanglement revival when $g_1/g_2\to 1$, which is the border between the topological and trivial phases, strongly suggests that the entanglement revival is closely connected with the edge mode.

Finally, while we have exclusively discussed the logarithmic negativity in this section, the universal revival can also be found in the mutual information (see Appendix~\ref{Appendix:A}). 
The von Neumann and R\'enyi entropy densities exhibit a size-independent behavior as well (see Appnendix~\ref{Appendix:B}), suggesting that the topological properties of the system affect the dynamical evolution and robustness of local correlations, both quantum and classical, in a universal way.

\section{Conclusion}
\label{Sec: Summary}

In this study of the fermionic and bosonic SSH models, characterized by the presence of a chiral symmetry, we have explored the dynamics of entanglement of a single site with the rest of the chain, in the presence of decoherence which preserves the chiral symmetry.
Our results reveal a remarkable entanglement revival which, for a fixed coupling ratio, acquires a universal profile when the system length exceeds the localization length of the zero-energy edge modes. 
The effective length scale for the onset of universality is identified with the localization length of the zero-energy edge modes.
By revealing a direct connection between the universal entanglement revival and the topological phase, these findings enhance our comprehension of the impact of topological features on entanglement dynamics and suggest promising applications in adaptable quantum information processing systems, including photonic quantum computers and semiconductor quantum dot arrays.
These results not only highlight the influence of topological features on entanglement dynamics but also motivate further investigation into various other decoherence scenarios. In particular, the effect of colored noise with chiral symmetry, which is relevant in experimental situations, is another interesting open question for future studies.

\section*{Acknowledgments}

C.D.\ acknowledges the support from the STU Scientific Research Initiation Grant (Grant No.~NTF25003T).
C.D.\ and M.-S.C.\ have been supported by the National Research Function (NRF) of Korea
(Grant Nos.~2022M3H3A106307411, 2023R1A2C1005588, RS-2024-00432563, RS-2024-00404854) and by the Ministry of Education through the BK21 Four program.
S.C.\ acknowledges support from the Innovation Program for Quantum Science and Technology (Grant No.~2021ZD0301602) and the National Science Association Funds (Grant No.~U2230402).

\appendix

\section{Mutual information}
\label{Appendix:A}
\renewcommand{\thefigure}{A\arabic{figure}}
\setcounter{figure}{0}

Mutual information is a measure of total correlations, both classical and quantum, between two subsystems. It provides a different perspective on the state of the system than entanglement measures such as logarithmic negativity. 
Furthermore, mutual information is usually easier to compute than most entanglement measures, making it a practical alternative for larger systems.
Here, we show that mutual information, like logarithmic negativity, exhibits a universal revival behavior, which indicates that universality is not limited to a specific entanglement measure.

The mutual information between subsystems $\mathbb{A}$ and $\mathbb{B}$ is defined by
\begin{equation}
\mathrm{I}_{\gamma}(\mathbb{A},\mathbb{B}) 
:= S_{\gamma}(\hat\rho_\mathbb{A}) + S_{\gamma}(\hat\rho_\mathbb{B}) - S_{\gamma}(\hat\rho),
\end{equation}
where $\hat\rho_{\mathbb{A}/\mathbb{B}}$ is the reduced density matrix of subsystem $\mathbb{A}/\mathbb{B}$ and $ \hat\rho$ the full density matrix, while the R\'enyi entropy of order $\gamma$ is defined as:
\begin{equation}\label{Sgamma}
S_{\gamma}(\hat\rho) := \frac{1}{1-\gamma} \log_2 \mathrm{Tr}(\hat\rho^\gamma).
\end{equation} 
In the limit $\gamma\rightarrow 1$, the R\'enyi entropy becomes the von Neumann entropy,
$S_{\mathrm{vN}}(\hat\rho) = -\mathrm{Tr}(\hat\rho \log_2 \hat\rho)$. 
Note that computing the von Neumann entropy (i.e., the R\'enyi entropy of order 1) may be computationally intensive for large systems. 
On the other hand, the R\'enyi entropy of order 2,
\begin{math}
S_2(\hat\rho) = -\log_2 \mathrm{Tr}(\hat\rho^2), 
\end{math}
is much easier to compute and advantageous for large or complex systems.

\begin{figure}
\centering
\includegraphics[width=80mm]{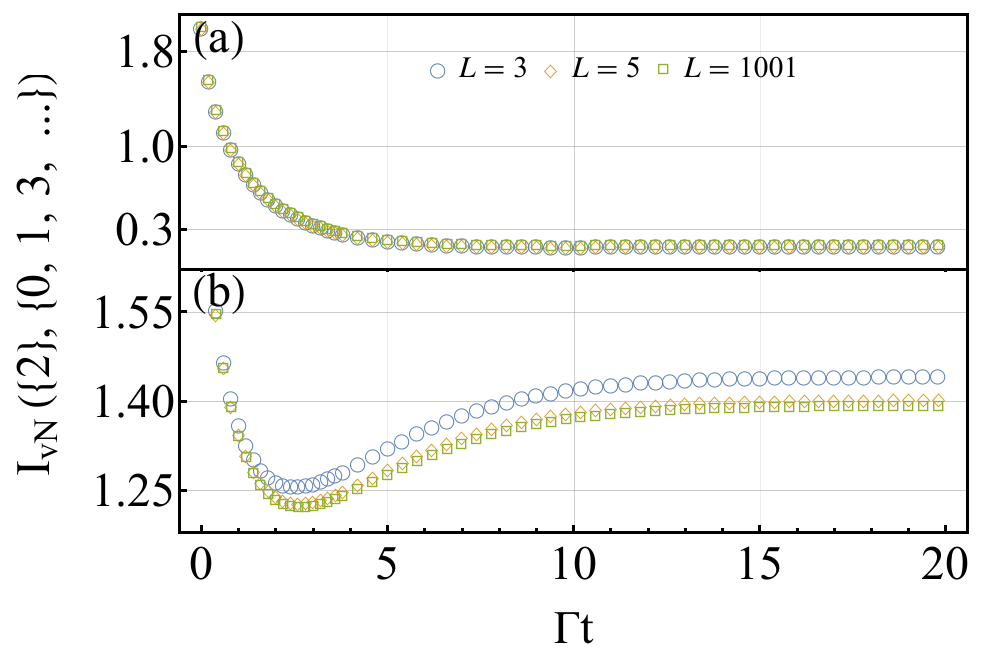}
\caption{Time dependence of the von Neumann mutual information
$\mathrm{I}_{\mathrm{vN}}(\{2\},\{0,1,3,...\})$ between site 2 and the rest of the chain in the fermionic SSH model for (a) $g_1/g_2=0.1$ and (b) $g_1/g_2=0.5$. In both panels, $\Gamma/g_2=0.1$.}
\label{Fig:MIfermion}
\end{figure}

For fermionic Gaussian states, the von Neumann mutual information can be calculated efficiently from the correlation function~\cite{Casini2009_JPA}. 
In Fig.~\ref{Fig:MIfermion}, we observe that the revival behavior disappears in certain limits, such as $g_1 \ll g_2$, because the mutual information is not a proper measure of quantum entanglement. 
However, the universal behavior always persists, which is in agreement with the results for the logarithmic negativity presented in the main text. 
In addition, both measures suggest that, for a fixed coupling ratio $g_1/g_2$, the universal behavior can be observed when the size of the half-chain SSH model exceeds a certain crossover length. 
The consistency in behavior between the two measures underscores the robustness of the universal entanglement revival phenomenon, confirming once more that the universal behavior is determined by the inherent properties of the system.

\begin{figure}
\centering
\includegraphics[width=80mm]{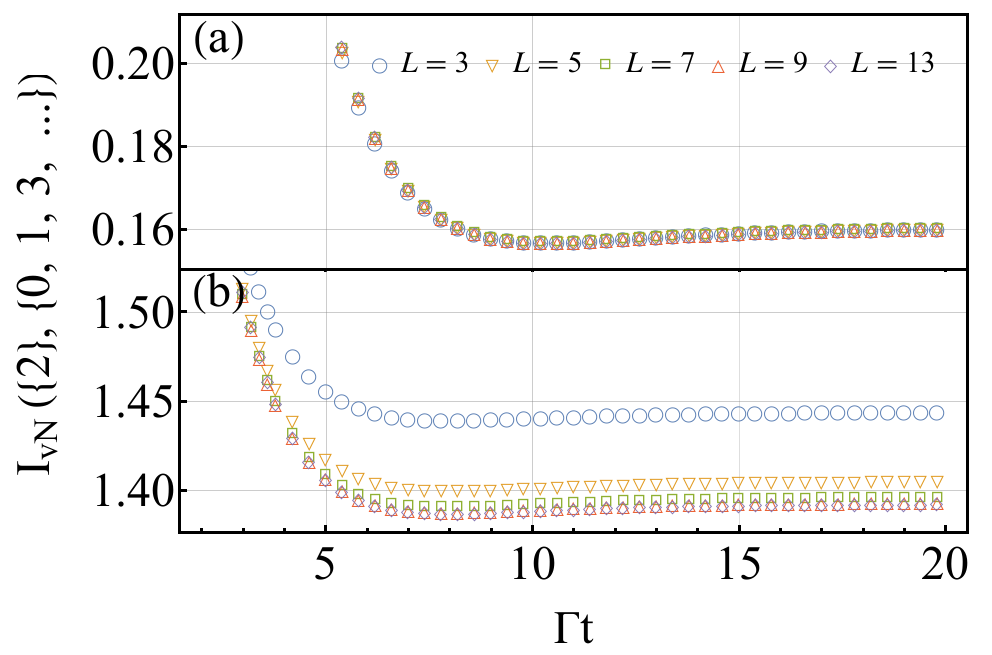}
\caption{The revival behavior of von Neumann mutual information in the bosonic model for coupling ratios for (a) $g_1/g_2=0.1$ and (b) $g_1/g_2=0.5$. In both panels, $\Gamma/g_2=0.1$.}
\label{Fig:MIboson}
\end{figure}

For the bosonic case, computing the von Neumann mutual information is challenging for large systems since the Hilbert space grows very fast with the system size and the relevant states are not Gaussian.
Currently, we are able to calculate the von Neumann mutual information up to $L=13$. Figure~\ref{Fig:MIboson} illustrates the entanglement dynamics for bosonic systems with different coupling ratios and chain lengths. 
As seen, despite the maximum length being much shorter than the fermion case, the universal behavior is still evident. 
Thus, the conclusions for bosons are very similar to those for fermions in Fig.~\ref{Fig:MIfermion}, and to the behavior of the logarithmic negativity.

\begin{figure}[tb]
\centering
\includegraphics[width=80mm]{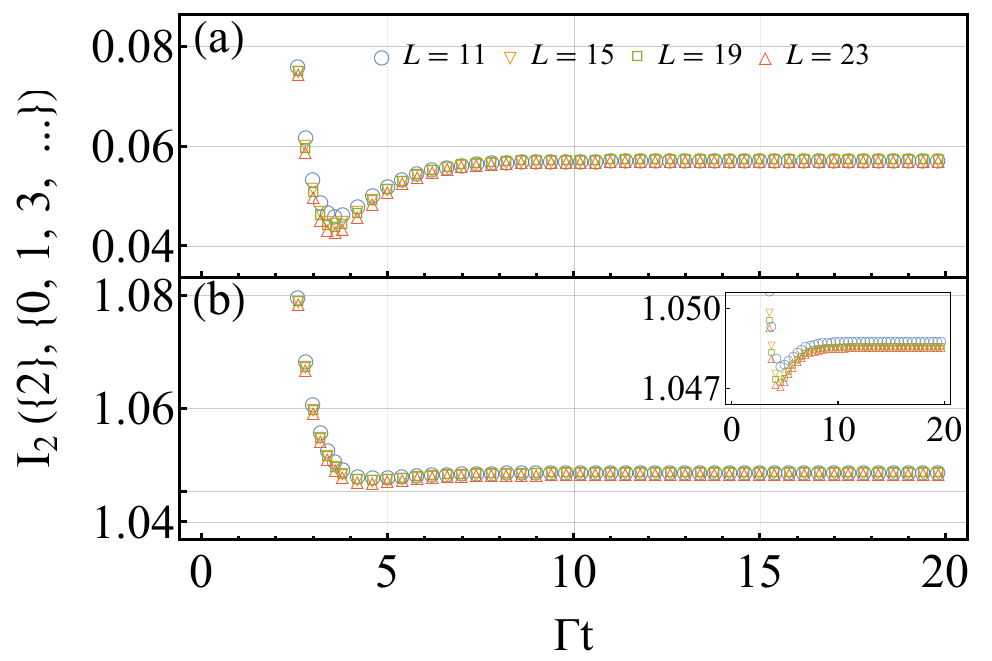}
\caption{Revival behavior of the R\'enyi mutual information of order 2 in the bosonic model for coupling ratios (a) $g_1/g_2=0.1$ and (b) $g_1/g_2=0.5$. 
Shown is the R\'enyi mutual information between site 2 and the rest of the chain. The inset of panel (b) is a zoom-in of the main plot. 
We have used $\Gamma/g_2=0.1$.}
\label{Fig:RenyiMIboson}
\end{figure}

Due to the limitations in calculating the von Neumann mutual information for bosonic systems, we extended our analysis to the order-2 R\'enyi mutual information, which can be obtained for larger system sizes (specifically, we consider $L = 11, 15, 19$, and $23$). 
As shown in Fig.~\ref{Fig:RenyiMIboson}, the R\'enyi mutual information also displays a universal revival. The larger accessible system sizes allow us to obtain good convergence also in Fig.~\ref{Fig:RenyiMIboson}(b), where a larger coupling ratio $g_1/g_2=0.5$ is assumed. 
This extension further confirms that the universal behavior is not restricted to the von Neumann mutual information and logarithmic negativity, offering new evidence of its robustness in larger bosonic systems.

\section{Entropy density}
\label{Appendix:B}

We also consider the R\'enyi entropy per site, $s_\gamma(\hat\rho)= S_\gamma(\hat \rho)/n$, where $S_\gamma(\hat \rho)$ is defined in Eq.~\eqref{Sgamma} and $n=(L-1)/2$ is the number of two-site unit cells. 
This quantity can be easily evaluated within the semiclassical approximation in  Eqs.~(\ref{Eq:rho}) and (\ref{Eq:probability}). 
In particular, the von Neumann entropy density ($\gamma\to 1$), $s_{\text{vN}}(\rho)$, is given by
\begin{equation}
\label{svN}
s_{\mathrm{vN}}=-\log_2 (1 - e^{-\Gamma t/2}) + e^{-\Gamma t/2} \log_2 (-1 + e^{\Gamma t/2}),
\end{equation}
and the R\'enyi entropy density of order 2 reads as
\begin{equation}
\label{sRenyi}
s_2=-\log_2 (1 + 2e^{-\Gamma t} - 2e^{-\Gamma t/2}).
\end{equation}
The qualitative time dependence of these quantities features dynamical behavior distinct from that of entanglement measures such as logarithmic negativity.
Since the system is in a pure state at $t=0$ and $t\to \infty$, they vanish at $t=0$ and $t=\infty$. Instead, the entropy density attains its maximum value $s_{\mathrm{vN}}=s_{2}=1$ at $\Gamma t=2\ln 2$. 
The increase of the entropy at intermediate times is in qualitative agreement with the reduction of purity and the occurrence of a minimum in the logarithmic negativity. 
Interestingly, in contrast to entanglement, the maximum entropy always occurs at $\Gamma t=2\ln 2$, independent of the coupling strengths $g_{1,2}$. 
This difference is due to the fact that the probabilities of the eigenstates, given in Eq.~(\ref{Eq:probability}), do not depend on $g_{1,2}$. 
Instead, the entanglement is obviously affected by the specific eigenstates of $\hat \rho$, thus the position of the minimum depends the coupling ratio $g_1/g_2$. 
Despite these differences, we note that both Eqs.~(\ref{svN}) and (\ref{sRenyi}) are independent of $L$, i.e., they also display a universal behavior, demonstrating that local quantum correlations are robust with respect to the overall system size. 
These results suggest that the robustness of local quantum correlations is likely due to the topological properties of the SSH model, contributing to the stability of the entanglement revival under different conditions.

\section{Spatially correlated noise}
\label{Appendix:C}

\begin{figure}
\includegraphics[width=80mm]{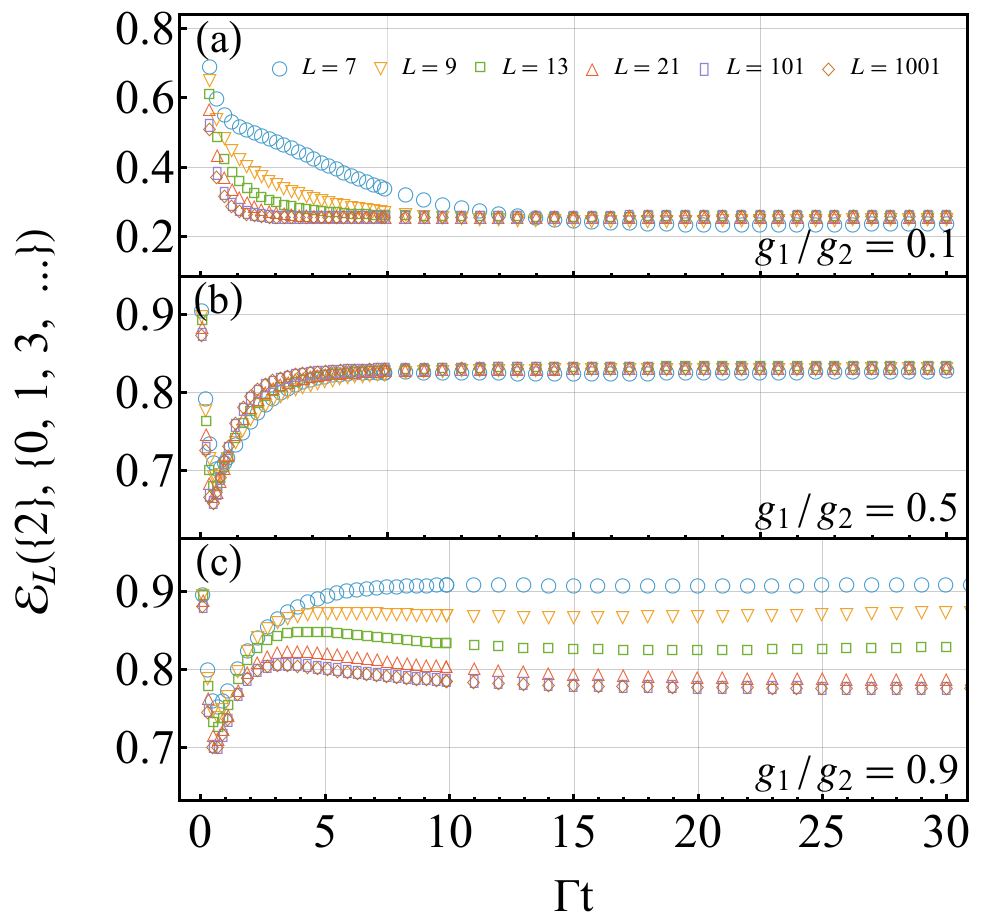} 
\caption{Entanglement dynamics under nearest-neighbor correlated decoherence, $\hat L_i=\hat f_{2i-1}+ \hat f_{2i+1}$. The entanglement is shown for various system sizes, ranging from $L = 7$ to $L = 1001$, and under different coupling ratios: (a) $g_1/g_2 = 0.1$, (b) $g_1/g_2 = 0.5$, and (c) $g_1/g_2 = 0.9$. The parameter that we take: $g_2=1, \Gamma/g_2=0.1$}
\label{Fig:correlatednoise1}
\end{figure}

\begin{figure}
\includegraphics[width=80mm]{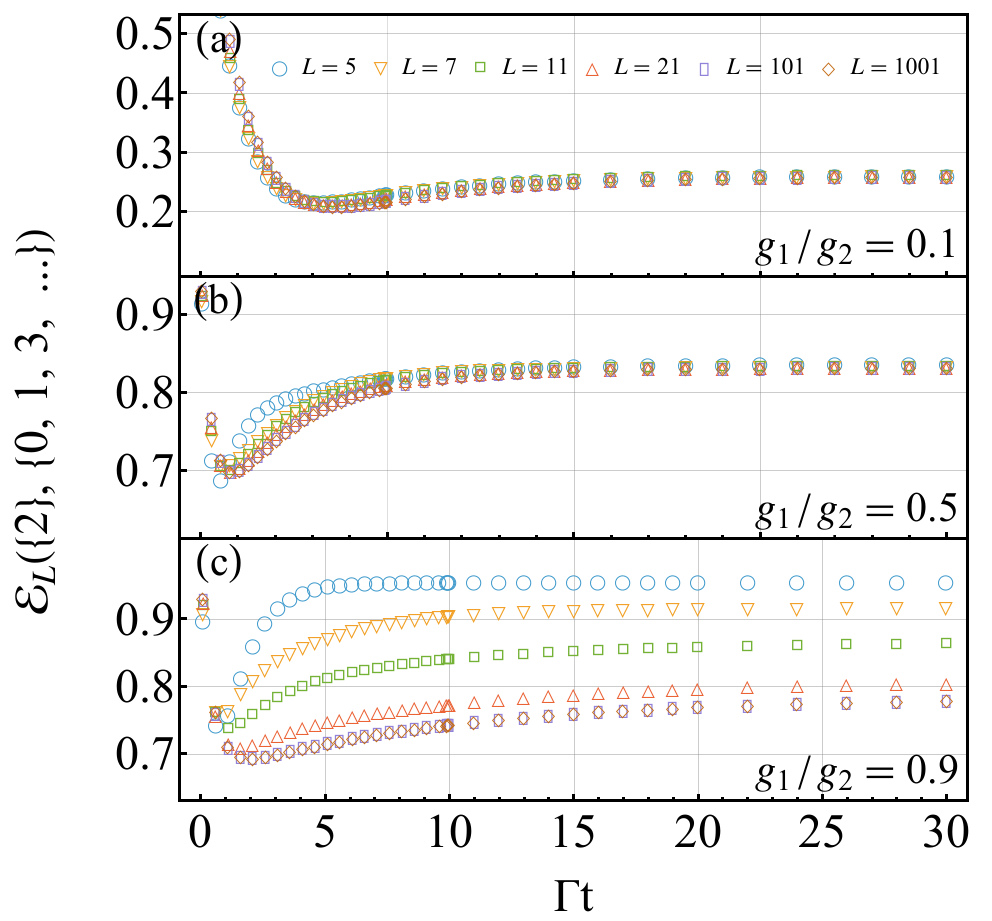} 
\caption{Entanglement dynamics under exponentially decaying correlated decoherence, $\hat L_i = \hat f_{2i-1} + e^{-1}(\hat f_{2i+1} + \hat f_{2i-3}) + e^{-2}(\hat f_{2i+3} + \hat f_{2i-5}) + \dots$. The entanglement is shown for various system sizes, ranging from $L = 5$ to $L = 1001$, and under different coupling ratios: (a) $g_1/g_2 = 0.1$, (b) $g_1/g_2 = 0.5$, and (c) $g_1/g_2 = 0.9$. The parameter that we take: $g_2=1, \Gamma/g_2=0.1$.}
\label{Fig:correlatednoise2}
\end{figure}

In the main text we assumed for simplicity a strictly local  decoherence channel. In realistic platforms, however, spatially correlated noise may arise and influence the dynamics. Inter-qubit noise correlations have been experimentally identified in densely packed silicon spin qubits~\cite{Yoneda2023_NaturePhysics}, and their effects on qubit dynamics have been theoretically analyzed in both Markovian and non-Markovian regimes~\cite{Zou2024_npj}. These studies underscore the importance of considering correlated noise when investigating entanglement under decoherence. 

To test the robustness of our results, we consider a decay channel with spatial correlations, which nevertheless remain short-ranged. This distinction is crucial, as topological properties are robust against local fluctuations and disorder, but can break down under globally correlated perturbations. For definiteness, here we restrict ourselves to the fermionic case and examine two representative types of correlated Lindblad operators that preserve the chiral symmetry of the SSH model: (i) The nearest-neighbor symmetric combination $\hat L_i=\hat f_{2i-1}+\hat f_{2i+1}$, where decoherence acts collectively on nearest-neighbor odd sites; (ii) The exponentially decaying form $\hat L_i = \hat f_{2i-1} + e^{-1}(\hat f_{2i+1} + \hat f_{2i-3}) + e^{-2}(\hat f_{2i+3} + \hat f_{2i-5}) + \dots$, which introduces a broader spatial profile of correlated decoherence, mimicking decaying interactions often present in realistic environments.

As shown in Fig.~\ref{Fig:correlatednoise1} and~\ref{Fig:correlatednoise2}, the entanglement dynamics under such spatially correlated decoherence remains qualitatively similar to that observed with local noise (cf. Fig.~\ref{Fig:universalrevival}).  The universal entanglement revival persists across various coupling ratios and becomes more pronounced in larger systems.  
We note that for $g_1/g_2 = 0.9$, the universal behavior is less visible in \emph{small} systems [see Fig.~\ref{Fig:correlatednoise1} (c) and~\ref{Fig:correlatednoise2} (c)] since the system size must exceed the localization length of edge mode, which grows fast as $g_1/g_2$ approaches 1.
All these results demonstrate that the universal entanglement revival persists for spatially correlated noise, as long as it preserves the chiral symmetry and the correlation length is not too long compared with the system size. In fact, as already noted, topological states are protected against local perturbations, but not global ones.

\section{Finite temperature}
\label{Appendix:D}

\begin{figure}
\includegraphics[width=80mm]{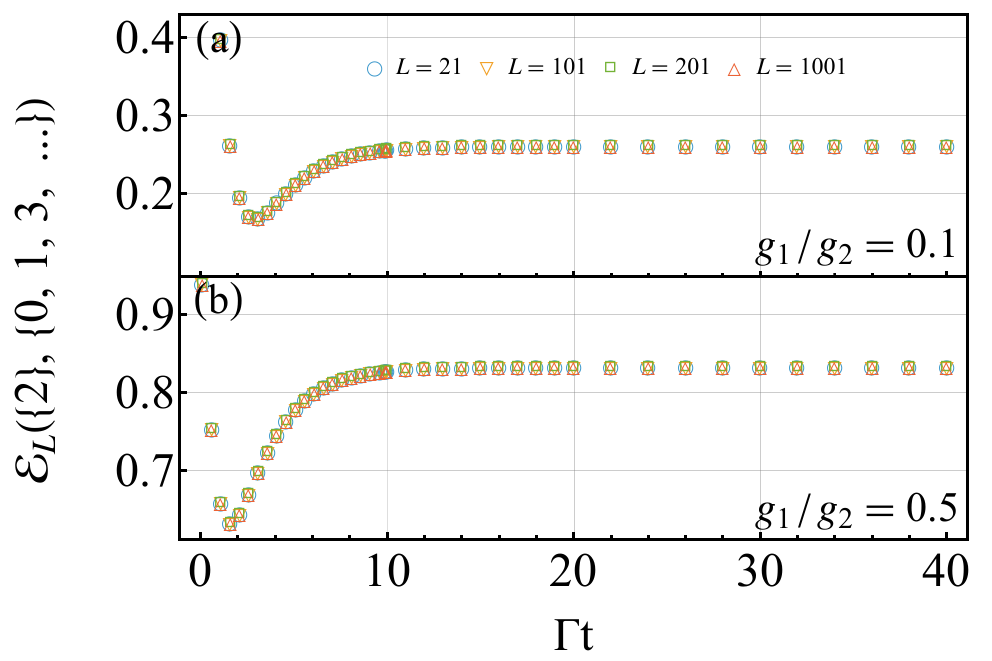} 
\caption{Entanglement dynamics, assuming that the small fraction $\delta = 0.05$ of negative-energy modes with the highest energy is left unoccupied, while all deeper modes and the edge mode remain filled. Explicitly, the initial state is $|\psi_g\rangle = \hat{c}_0^\dagger \prod_{k=1}^{(L-1)/2-\delta (L-1)} \hat{c}_{-k}^\dagger \ket{\ }$. We take $\Gamma/g_2=0.1$.}
\label{Fig:finitetemp1}
\end{figure}

\begin{figure}
\includegraphics[width=80mm]{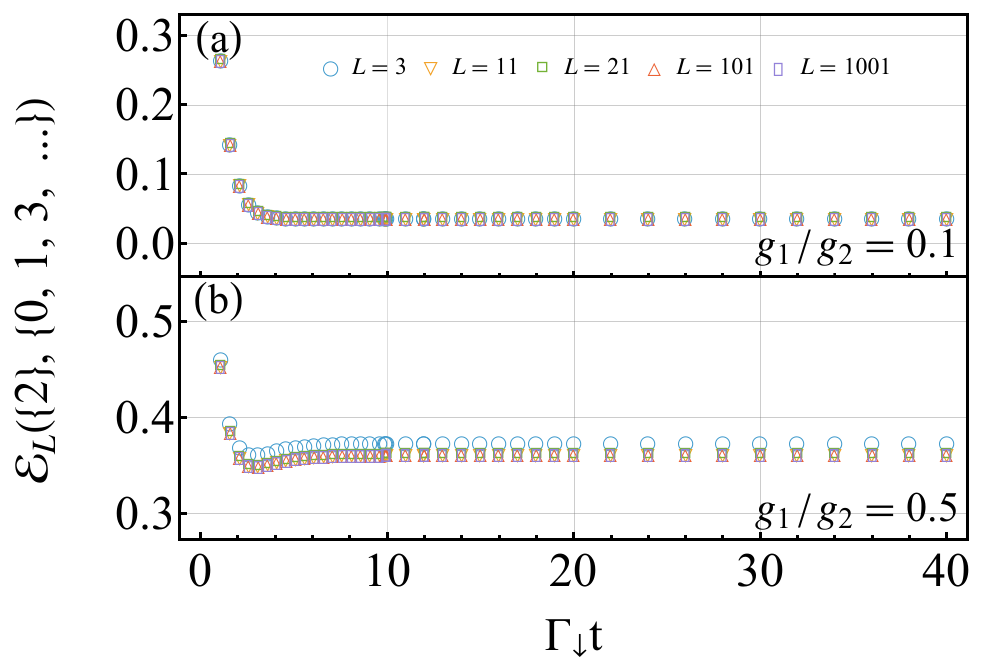} 
\caption{Entanglement dynamics including thermal activation processes, simulated by adding Lindblad operators $\hat{L}_i = \hat{f}_{2i-1}^\dagger$. The ratio between the excitation and decay rates is set to $\Gamma_{\uparrow} / \Gamma_{\downarrow} = 0.5$, and $\Gamma_{\downarrow} / g_2= 0.1$.
}
\label{Fig:finitetemp2}
\end{figure}

In realistic quantum platforms, finite temperature is unavoidable. It may affect the preparation of the initial state, as well as the subsequent dynamics. To examine whether the entanglement revival behavior persists under such conditions, we consider two main effects: (i) thermal excitations in the initial state, and (ii) thermal activation processes during the dynamics.

In the first case, we suppose that the temperature $T$ satisfies $T \ll g_{1,2}$, and that the chemical potential $\mu$ is close to the upper edge of the negative-energy band. From Eq.~(\ref{Eq: energy}), the latter condition is easily written as $\mu \approx -g_2+g_1$, and is realized when $k \approx (L-1)/2$. Since the energy gap is $2(g_2-g_1)$, the population of the positive-energy states remains negligible (at least, sufficiency far from the boundary of the topological phase). It follows that the main consequence of a finite temperature is the depopulation of energy states close to the chemical potential, while deeper negative-energy modes and the zero-energy edge mode remain occupied. To capture this effect, we consider the initial state  $|\psi_g\rangle=\hat{c}_0^\dagger \prod_{k=1}^{(L-1)/2 - \delta(L-1)} \hat{c}_{-k}^\dagger\,\ket{\ }$, where $\delta$ indicates the small fraction of unoccupied negative-energy states. As shown in Fig.~\ref{Fig:finitetemp1}, the universal entanglement revival behavior exhibits excellent robustness. 

Although the above result has been obtained by assuming, for simplicity, a pure initial state, we should also stress that the system undergoes a dramatic loss of purity in the course of the time evolution, due to the simultaneous decay of all occupied modes. Therefore, even for a thermally mixed initial state, we do not expect significant qualitative changes to the revival dynamics in the low-temperature regime $T \ll g_{1,2}$. The main difference will be a small decrease of purity (and entanglement) at the beginning of the time evolution.    

We next consider the effect of thermal activation, which may induce tunneling  processes from the external reservoirs into the SSH chain. Correspondingly, we include additional Lindblad operators of the form $\hat L_i =\hat f_{2i-1}^\dagger$ in the master equation:
\begin{equation}
\frac{d\hat \rho}{dt} = -i[\hat H,\hat \rho]
+\sum_{i=1}^{(L-1)/2}( \Gamma_{\downarrow} \mathcal{D}[\hat f_{2i-1}]\hat \rho+\Gamma_{\uparrow} \mathcal{D}[\hat f_{2i-1}^\dagger]\hat \rho),
\end{equation}
where  $\Gamma_{\uparrow}  < \Gamma_{\downarrow} $. As shown in Fig.~\ref{Fig:finitetemp2}, the universal behavior remains clearly visible at a relatively large ratio $\Gamma_{\uparrow}  / \Gamma_{\downarrow} =0.5$, which demonstrates the robustness of the mechanism in the presence of thermal noise. Nevertheless, since more excited states are involved in the dynamics, the entanglement values and the visibility of the revival are moderately reduced compared to the zero-temperature case, particularly at $g_1/g_2 = 0.1$.


\end{document}